\documentclass[]{elsarticle}

\usepackage[utf8]{inputenc}
\usepackage[T1]{fontenc}
\usepackage{graphicx}
\usepackage{dblfloatfix}
\usepackage{upgreek}
\usepackage[colorlinks]{hyperref}

\newcommand{\ohm}{$\Upomega$}
\newcommand{\sref}[1]{\hyperref[suppl]{#1}}

\makeatletter
\providecommand{\doi}[1]{%
  \begingroup
    \let\bibinfo\@secondoftwo
    \urlstyle{rm}%
    \href{http://dx.doi.org/#1}{%
      doi:\discretionary{}{}{}%
      \nolinkurl{#1}%
    }%
  \endgroup
}
\makeatother

\begin{document}

\begin{frontmatter}

    \title{Effect of substrate temperature on the deposition of Al-Doped ZnO thin films using high power impulse magnetron sputtering}

    \author[ul,liu]{Martin Mickan}
    \author[liu]{Ulf Helmersson}
    \author[ul]{David Horwat\corref{cor1}}
    \ead{david.horwat@univ-lorraine.fr}

    \address[ul]{Université de Lorraine, Institut Jean Lamour, UMR7198, Nancy, F-54011, France}
    \address[liu]{Plasma \& Coatings Physics Division, IFM–Material Physics, Linköping University, SE-581 83 Linköping, Sweden}

    \cortext[cor1]{Corresponding author}

    \begin{abstract}
        Al-doped ZnO thin films were deposited using reactive high power impulse magnetron sputtering at substrate temperatures between room temperature and 600\,$^\circ$C.
        Two sample series with different oxygen partial pressures were studied.
        The films with the lowest resistivity of $3 \times 10^{-4}$\,\ohm cm were deposited at the highest substrate temperature of 600\,$^\circ$C.
        The improvement of the electrical properties could be related to an improvement of the mobility due to the improved crystallinity.
        This improved crystallinity also increased the stability of the films towards ambient moisture.
        On the other hand, the detrimental influence of negative oxygen bombardment could be avoided, as the HiPIMS process can take place in the metal or transition mode even at relatively high oxygen partial pressures.
    \end{abstract}


\end{frontmatter}

\section{Introduction}
Transparent conductive oxides (TCOs) are important materials for wide ranges of applications in optoelectronics and energy conversion, that can be used for example as transparent electrodes in solar cells~\cite{beyer_transparent_2007} or flat panel displays~\cite{lan_alternative_1995}.
A common TCO is ZnO, that is usually doped with group III elements such as Al, in order to achieve low resistivity.
However, the solubility limit of Al in ZnO is quite low at around 0.3\,at\%~\cite{shirouzu_distribution_2007}.
To reach high active dopant concentrations in Al doped ZnO (AZO) films, the films need to be grown using a non-equilibrium method such as physical vapor deposition (PVD)~\cite{zakutayev_non-equilibrium_2013}.
A commonly used  technique for the deposition of AZO films is direct current (DC) or radio frequency (RF)  magnetron sputtering as it can be easily scaled up to large deposition areas~\cite{cornelius_achieving_2009}.

As the properties of the growing films are governed by both kinetics and thermodynamics, the substrate temperature is an important parameter in the deposition of AZO films.
There have been several studies about the influence of the substrate temperature in magnetron sputter deposition of AZO films in the case of DC and RF magnetron sputtering.
\citet{he_effect_2013} studied DC sputtering of AZO films at deposition temperatures up to 480\,$^\circ$C.
They observed an increase in both mobility and charge carrier concentration with increasing substrate temperature.
Similarly, \citet{park_deposition-temperature_2006} observed an improvement of the electrical properties with  the substrate temperature in their study of RF magnetron sputtering of AZO films at substrate temperatures up to the 500\,$^\circ$C.
On the other hand, \citet{vinnichenko_establishing_2010} observed a degradation of the electrical properties above an optimum temperature of 200\,$^\circ$C.
They explain this by the formation of an insulating metastable phase at higher temperatures.
\citet{bikowski_influence_2012} find an optimum temperature of 300\,$^\circ$C in the case of RF magnetron sputtering of AZO films.

An upcoming technique is high power impulse magnetron sputtering (HiPIMS).
HiPIMS uses short pulses of high current density~\cite{kouznetsov_novel_1999}.
Recently, it has been shown that highly conductive and transparent AZO films can be deposited without substrate heating using HiPIMS~\cite{mickan_room_2016}.
However, there are only few studies that investigate the influence of the substrate temperature in the case of HiPIMS deposition of AZO films.
\citet{ruske_reactive_2008} used substrate temperatures of up to 200\,$^\circ$C and improved the resistivity of the AZO films by an order of magnitude as compared to deposition at room temperature.

There seems to be no clear conclusion about which temperature is the optimum temperature for the deposition of AZO thin films.
The optimum temperature might depend on the deposition technique, as well as the deposition parameters.
In this work, AZO films are deposited by reactive HiPIMS at substrate temperatures between room temperature and 600\,$^\circ$C using two different O$_2$/Ar ratios.
Their electrical, optical and structural properties are investigated.
Another common effect in AZO films is the degradation of the electrical properties in a humid environment~\cite{mickan_restoring_2017}.
It has been shown that a heat treatment of deposited AZO films at 650\,$^\circ$C in vacuum can make the films stable towards humid environments~\cite{hupkes_damp_2014}.
Therefore, the aging behaviour of the films in ambient moisture is also studied.

\section{Experimental}
AZO films were deposited onto Si and fused silica substrates mounted on a rotatable substrate holder using reactive HiPIMS of a 50\,mm Zn/Al target with 2\,at\% of Al.
The substrate temperature was controlled with a resistive heater for two different sample series.
In series~1, the temperature of the substrate was varied between room temperature and 500\,$^\circ$C and in series~2 the temperature was varied between room temperature and 600\,$^\circ$C.
The substrate temperature without intentional heating is expected not to exceed 45\,$^\circ$C~\cite{mickan_room_2016}.
The oxygen and argon flow rates have been slightly changed between the two series as indicated in table~\ref{tab:experimental} in order to assess the robustness of the process.
The total pressure in the chamber was controlled by a throttle valve to be 1\,Pa in all cases.
The HiPIMS unit (Melec Spik 2000A) was charged by an Advanced Energy Pinnacle Plus DC power supply.
The DC supply was set to a constant voltage of 560\,V and the HiPIMS unit delivered pulses of a length of 100\,$\upmu$s with a frequency of 1000\,Hz.
This leads to an average power of around 120\,W and a peak current of around 2.6\,A.
The distance between the target and the substrate was 11\,cm and the deposition time was 10\,minutes.

\begin{table}
    \centering
    \caption{\label{tab:experimental}Experimental conditions for the two sample series}
    \begin{tabular}{lccc}
        \hline
        & \shortstack{Ar flow rate \\ (sccm)} & \shortstack{O$_2$ flow rate \\ (sccm)} & \shortstack{O$_2$/Ar ratio\\(\%)} \\
        \hline
        Series~1 & 47 & 17 & 36.2 \\
        Series~2 & 59 & 24 & 40.7 \\
        \hline
    \end{tabular}
\end{table}

The electrical properties of the films were measured using a HMS 5000 Hall effect  measurement setup in the van der Pauw geometry~\cite{van_der_pauw_l.j._method_1958}.
Transmittance measurements were performed using a Cary 5000 UV-Vis-NIR spectrophotometer.
The thickness of the films was measured using cross-sectional back-scattered electron images in a Philips XL30 scanning electron microscope (SEM) with an acceleration voltage of 15\,kV.
Additional images were taken with a LEO 1550 Gemini SEM with an acceleration voltage of 5\,kV.
A rough estimation of the Al concentration in the films was obtained by energy dispersive spectrometry (EDS) in the LEO 1550 Gemini SEM on selected samples.
In order to reduce the influence of the substrate on the EDS measurement, the samples were tilted by 30\,$^\circ$ and an acceleration voltage of 3\,kV was used.
Crystallographic information was obtained using $\theta/2\theta$ x-ray diffraction (XRD).
Two different diffractometers with Cu anodes were used.
An AXS Bruker D8 Advance with an automatic sample loading system was used to scan a large $2\theta$ range from 10 to 120\,$^\circ$.
Scans in the range of 30 to 40\,$^\circ$ were performed using a PANalytical X'pert diffractometer.
This allowed a fine-calibration of the alignment of the sample tilt to be able to compare the peak intensities between different samples.
In this case, the contribution of the Cu K$\alpha_2$ component was removed numerically using the method by \citet{rachinger_correction_1948}.
From the line profile of the (002) peak of ZnO, the microstrain and the size of the coherent scattering domains could be determined using the method by \citet{de_keijser_use_1982}.

\section{Results}
Magnetron sputtered AZO films often show a lateral variation of the properties along the substrate surface~\cite{jullien_influence_2011}.
It has been shown that this variation decreases in the case of HiPIMS deposited films~\cite{mickan_room_2016}.
There is also a lateral variation of the properties for the films in this experiment, however this lateral variation is much smaller than the variation of the properties with the temperature.
The lateral variation is therefore only shown in the errorbars in the figures.
The small lateral variation can be related both to the effects of the HiPIMS process, as well as the large distance between the target and the substrates.

\begin{figure}
    \centering
    \includegraphics[width=0.5\textwidth]{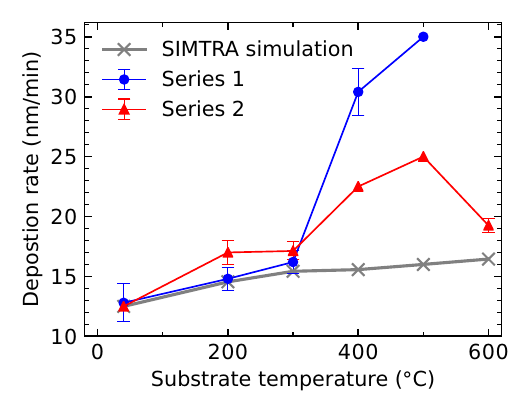}
    \caption{\label{fig:deposition-rate}Deposition rate of the AZO films as a function of temperature and the simulated normalized deposition rate using SIMTRA}
\end{figure}

Figure~\ref{fig:deposition-rate} shows the deposition rate as a function of the substrate temperature.
There is an 170\,\% (100\,\%) increase in deposition rate between room temperature and 500\,$^\circ$C for series~1(series~2).
The deposition rate decreases again at a temperature of 600\,$^\circ$C.
The increase in deposition rate at high temperature is stronger for sample series~1.
This result could have several possible reasons, such as the change in the gas density due to the temperature.
A change in gas density could have an effect on both the target condition as well as the film growth.

Under the assumption, that the gas temperature is in equilibrium with the substrate temperature, the gas density decreases by 60\,\% between room temperature and 500\,$^\circ$C, as the pressure is kept constant at 1\,Pa.
A decrease in the gas density causes an increase in the mean free path of the sputtered atoms, which could explain the increase in the deposition rate.
In order to estimate the magnitude of this effect, the simulated deposition rates using the software SIMTRA~\cite{depla_magnetron_2012} are also shown in figure~\ref{fig:deposition-rate}.
These simulations show only an increase of about 28\,\% in the deposition rate between room temperature and 500\,$^\circ$C, which is much lower than the increase observed in our films.

\begin{figure}
    \centering
    \includegraphics[width=0.5\textwidth]{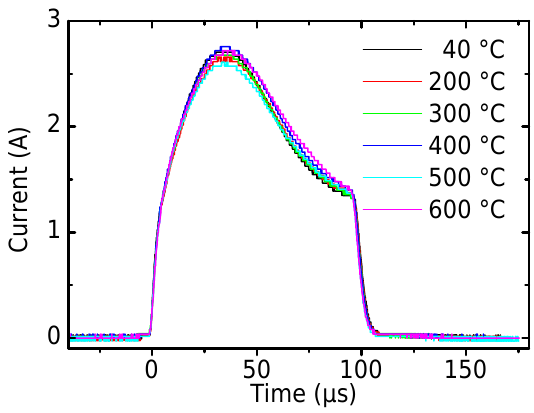}
    \caption{\label{fig:HiPIMS-current}Current waveforms of the depositions for sample series~2}
\end{figure}

Another possible explanation could be the decrease in the reactive gas density, that would move the process further into the metal mode and therefore increase the deposition rate.
The increase in deposition rate being more pronounced in series~1, i.e. at lower oxygen partial pressure, supports this interpretation.
A shift in the reactive sputter process towards the metal mode should affect the shape of the discharge current waveforms~\cite{aiempanakit_understanding_2013}.
However, the discharge current waveforms, shown in figure~\ref{fig:HiPIMS-current} for series~2, do not show any significant changes with the temperature.
For all the samples, the current reaches a peak value of about 2.7\,A after about 35\,$\upmu$s and decreases afterwards due to the rarefaction effect~\cite{horwat_compression_2010}.
The current waveforms for series~1 (shown in figure~\sref{S1} in the supplementary material) look similar.
There is no significant difference in the shape of the current waveforms and the peak current is relatively constant, suggesting that the effect of the substrate temperature on the target conditions should be small.

\begin{figure}
    \centering
    \includegraphics[width=0.5\textwidth]{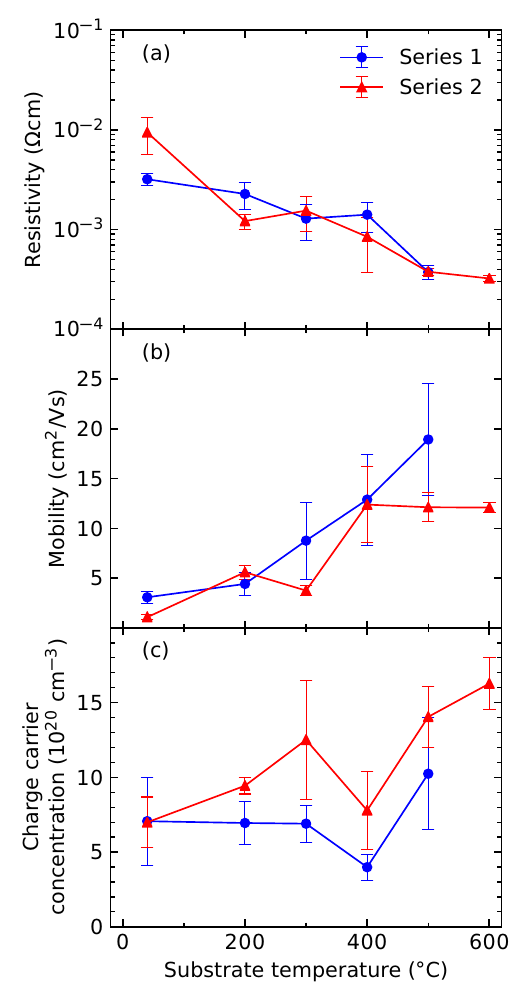}
    \caption{\label{fig:elec}Resistivity (a), mobility (b) and charge carrier concentration (c) of the AZO films as a function of substrate temperature}
\end{figure}

The resistivity of the films has been measured using the Hall effect setup.
The results are shown in figure~\ref{fig:elec}a as a function of substrate temperature.
The resistivity decreases with increasing substrate temperature for both series from a value of around $9 \times 10^{-3}$\,\ohm cm at room temperature to around $3 \times 10^{-4}$\,\ohm cm at 600\,$^\circ$C.
The evolution of the resistivity is very similar for both series despite the different deposition parameters.
This shows that the HiPIMS process indeed allows to obtain AZO films with good properties over a relatively large range of parameters.

The mobility and charge carrier concentration of the films have also been measured using the Hall effect setup, and are shown in figure~\ref{fig:elec}b and c, respectively.
The mobility is increasing with the substrate temperature for both sample series.
For the charge carrier concentration, no clear evolution with the substrate temperature is visible for series~1.
For series~2, the charge carrier concentration is increasing with the substrate temperature.
However, a dip in the charge carrier concentration is observed for the substrate temperature of 400\,$^\circ$C for both series obtained with the different sputtering conditions.

EDS measurements allowed to obtain a rough estimation of the chemical composition in the films.
The Al/ (Zn+Al) ratio is found to be between 2-3\,\% independently of the substrate temperature.
This means, that the increase in charge carrier concentration cannot be related to a change in the Al concentration.

\begin{figure}
    \centering
    \includegraphics[width=0.5\textwidth]{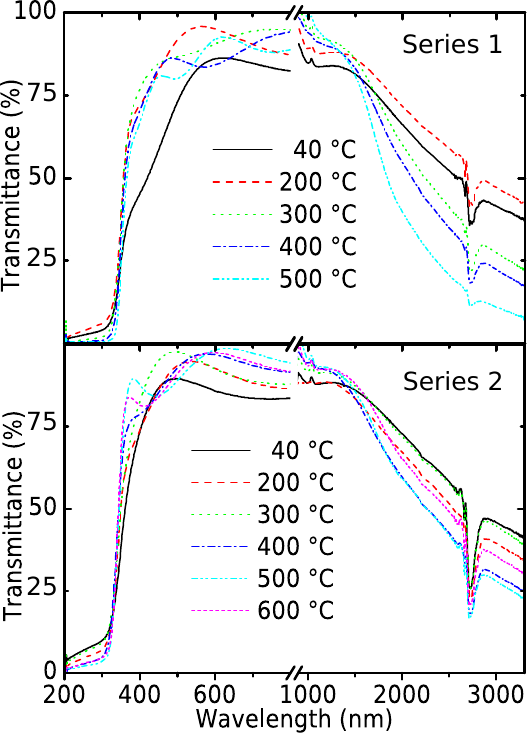}
    \caption{\label{fig:Transmittance-spectra}Transmittance spectra of AZO films from series~1 (top) and series~2 (bottom)}
\end{figure}

The transmittance of the films was measured in the spectrophotometer.
The resulting spectra for the samples of both series are shown in figure~\ref{fig:Transmittance-spectra} in both the visible and the near infrared range.
The transmittance in the near infrared range seems to be decreasing with the substrate temperature suggesting an increase in the charge carrier concentration corresponding to a shift in the plasmon resonance frequency towards lower wavelengths~\cite{chopra_transparent_1983}.

\begin{figure}
    \centering
    \includegraphics[width=0.5\textwidth]{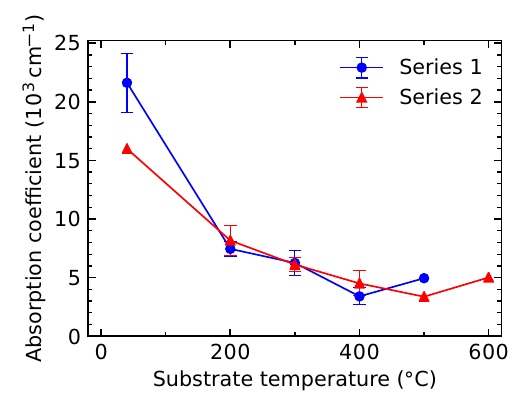}
    \caption{\label{fig:Absorption}Average absorption coefficient in the visible range as a function of substrate temperature}
\end{figure}

In order to avoid a biased interpretation of the transmittance spectra due to the variation of the film thickness, the absorption coefficient $\alpha$ was calculated as $\alpha=1/t \cdot \ln 1/T$, with $t$ the thickness of the films and $T$ the transmittance.
Figure~\ref{fig:Absorption} shows the average of $\alpha$ in the visible range as a function of the substrate temperature.
The absorption coefficient can be related to the level of oxygen substoichiometry of AZO as shown in several studies~\cite{jullien_influence_2011,horwat_effects_2007}.
Based on that, at room temperature, the film of series~1 could be more substoichiometric than the film of series~2.
Moreover, increasing the substrate temperature lowers the absolute value of the absorption coefficient and decreases the relative difference between the two series.
This could be explained by an increased chemical reactivity of oxygen at high temperatures under the assumption that the oxygen partial pressure is high enough to not limit the adsorption of oxygen on the film surface~\cite{hollands_mechanism_1968}.

\begin{figure}
    \centering
    \includegraphics[width=0.5\textwidth]{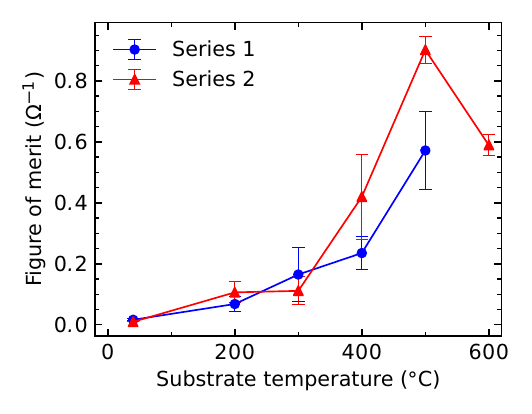}
    \caption{\label{fig:figure_of_merit}Figure of merit of the AZO films as a function of substrate temperature}
\end{figure}

To evaluate the overall performance of the films the figure of merit $\phi$ can be calculated from the absorption coefficient $\alpha$ and the resistivity $\rho$ as $\phi=1/(\alpha \cdot \rho)$ \cite{gordon_criteria_2000}.
The figure of merit for the two sample series is shown in figure~\ref{fig:figure_of_merit} as a function of the substrate temperature.
At room temperature, the figure of merit has a relatively low value 0.01\,\ohm$^{-1}$.
The maximum value of 0.9\,\ohm$^{-1}$ is found at 500\,$^\circ$C in case of series~2.
At 600\,$^\circ$C the figure of merit decreases again to a value of 0.6\,\ohm$^{-1}$.
In the literature the best values for the figure of merit of AZO films are about 5\,\ohm$^{-1}$ \cite{gordon_criteria_2000}.
\citet{fumagalli_room_2014} deposited AZO films using pulsed DC magnetron sputtering from a ceramic target and found a lateral variation of the figure of merit between 0.3 and 2.3\,\ohm$^{-1}$.
On the other hand, \citet{he_effect_2013} used reactive DC magnetron sputtering to deposit AZO films between room temperature and 480\,$^\circ$C.
From their reported values for the resistivity and the transmittance, the figure of merit can be calculated to increase from $4 \times 10^{-4}$\,\ohm$^{-1}$ to 0.3\,\ohm$^{-1}$ with the substrate temperature.

\begin{figure}
    \centering
    \includegraphics[width=0.5\textwidth]{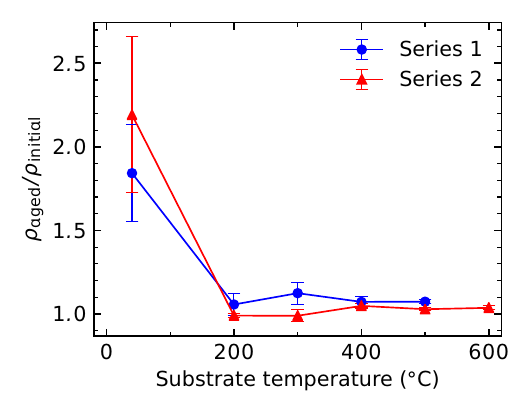}
    \caption{\label{fig:Resistivity-aged}Relative change in the resistivity as a function of substrate temperature after exposure of the films to ambient moisture for 9 months}
\end{figure}

The resistivity of our films has been measured again 9 months after the deposition in order to evaluate the properties of the films after exposure to ambient moisture with an average relative humidity around 35\,\%, similar to the aging procedure described elsewhere~\cite{mickan_restoring_2017}.
The relative change in the resistivity $(\rho_{\mathrm{aged}} / \rho_{\mathrm{initial}})$ is shown in figure~\ref{fig:Resistivity-aged}.
For the sample deposited at room temperature, the resistivity increases by 100\,\% after the aging.
For higher substrate temperatures, however, the resistivity does not change significantly over time.
Although the resistivity of AZO films is well known to degrade in contact with a humid environment, \citet{hupkes_damp_2014} have shown that deposition at 600\,$^\circ$C can prevent this effect.
They attributed this improvement to less defective grain boundaries.
Our results suggest that a substrate temperature as low as 200\,$^\circ$C can also be highly beneficial, although it is difficult to draw definitive conclusion from this result, as the aging of the films was performed at room temperature and not in a standardized damp heat treatment test.

The XRD patterns of the films in the range between 32 and 38\,$^\circ$ for series~1 are shown in figure~\ref{fig:XRD} and for series~2 in figure~\sref{S2} in the supplementary material.
The intensities are normalized according to the measured film thickness.
The XRD patterns show only a single peak around 34\,$^\circ$, which corresponds to the (002)  peak of the wurtzite structure of ZnO, which can also be seen in full range of the XRD patterns in figure~\sref{S3} in the supplementary material.
This means that the films are preferentially oriented along the c-axis.

\begin{figure}
    \centering
    \includegraphics[width=0.5\textwidth]{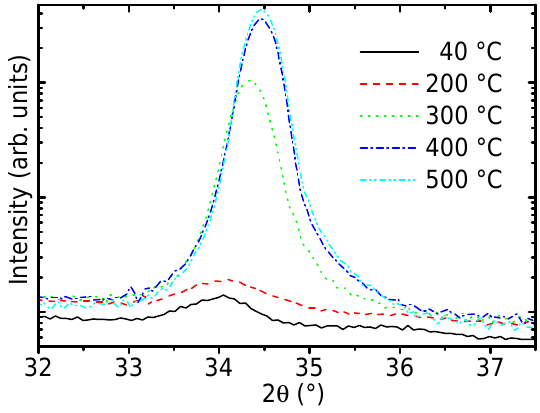}
    \caption{\label{fig:XRD}Normalized XRD patterns of the samples in series~1}
\end{figure}

\begin{figure}
    \centering
    \includegraphics[width=0.5\textwidth]{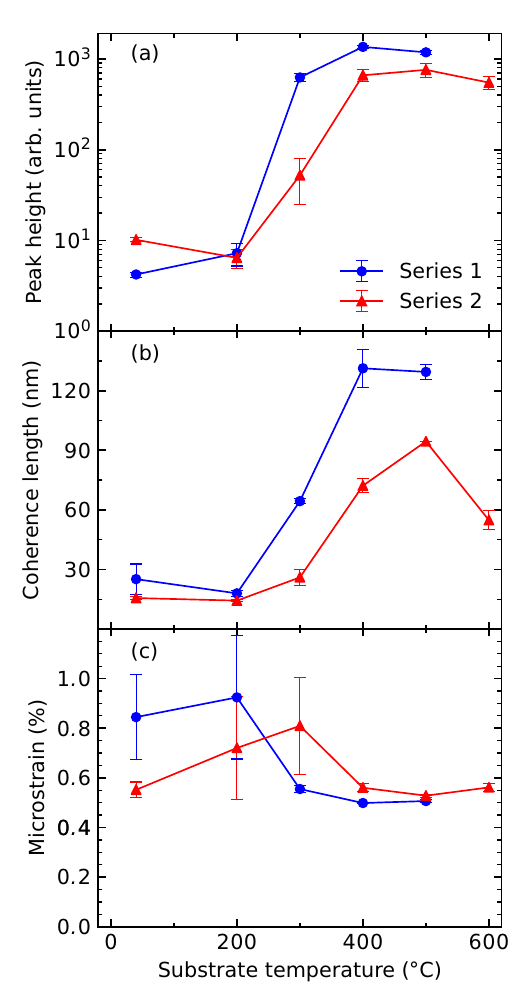}
    \caption{\label{fig:XRD-combined}Results from fitting the (002) peak in the XRD patterns as a function of substrate temperature: peak height (a), and coherence length (b) and microstrain (c), calculated from the Lorentzian and Gaussian contribution to the peak width, respectively.}
\end{figure}

The height of the (002) peak of the films is shown in figure~\ref{fig:XRD-combined}a as a function of substrate temperature.
This peak height is normalized with the film thickness and can therefore give an indication of the crystalline quality of the films.
The crystalline quality is increasing with the substrate temperature.
However, a saturation of the diffracted intensity is achieved at a substrate temperature of 400\,$^\circ$C and above.

Figure~\ref{fig:XRD-combined}b shows the size of coherent scattering domains calculated from the Lorentzian contribution to the width of the (002) peaks of the films as a function of the substrate temperature.
The coherence length is increasing with the substrate temperature, except for the highest temperature of 600\,$^\circ$C.
The evolution of the coherence length is similar to the evolution of the film thickness as seen in figure~\ref{fig:deposition-rate}.
However, the film thickness increases by a factor of 2-3 between room temperature and 500\,$^\circ$C, whereas the coherence length increases by a factor of 5-6.
This means that the increase in the coherence length cannot be attributed only to the increase in film thickness with the temperature.

The microstrain in the sample has also been calculated from the Gaussian contribution to the width of the (002) peaks and is shown in figure~\ref{fig:XRD-combined}c as a function of substrate temperature.
The microstrain is relatively low (around 0.8\,\%) for all the substrate temperatures and the variation is within the error.
\citet{bikowski_influence_2012} also did not observe any development of the microstrain with the substrate temperature in the case of RF sputtering of AZO films.
However, in their case the microstrain was lower than in our films.
On the other hand, \citet{cornelius_zno_2016} observed an increase in the microstrain with increasing substrate temperature for a target Al concentration of 1.7\,at\% in the case of DC sputtering, which they attributed to an increasing disorder due to an increase in the Al concentration in the film.

\begin{figure}
    \centering
    \includegraphics[width=0.5\textwidth]{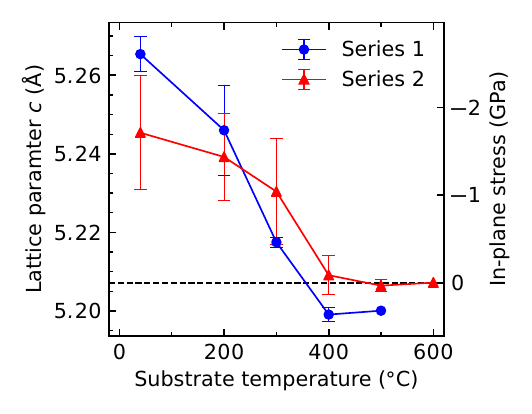}
    \caption{\label{fig:XRD-c}Lattice parameter $c$ calculated from the (002) peak. The horizontal line indicates the lattice parameter for bulk ZnO. The right scale bar shows the in-plane stress according to the biaxial strain model}
\end{figure}

From the position of the (002) peak the lattice parameter $c$ can be calculated, which is shown in figure~\ref{fig:XRD-c}.
Using the biaxial strain model, the in-plane stress can be calculated from the lattice parameter $c$ with
\begin{equation}
    \sigma = \frac{2 C_{13}^2 - C_{33} (C_{11} + C_{12})}{2 C_{13}} \cdot \frac{c-c_{\mathrm{bulk}}}{c_{\mathrm{bulk}}}
    \label{eqn:stress}
\end{equation}
where $c_{\mathrm{bulk}}=5.2071$\,\AA{} is the lattice parameter of bulk ZnO and $C_{ij}$ are the elastic constants of ZnO ($C_{11}=208.8$, $C_{33}=213.8$, $C_{12}=119.7$ and $C_{13}=104.2$\,GPa) \cite{cebulla_-doped_1998}.
At low temperatures the film is strained along the c-axis corresponding to a compressive in-plane stress.
For temperatures of 400\,$^\circ$C and above the c-axis parameter approaches the value of bulk ZnO meaning that the compressive stress is relieved, which can be related to the annealing of defects~\cite{cornelius_zno_2016}.
\citet{bikowski_influence_2012} observed a similar trend for deposition temperatures from 300\,$^\circ$C.
\citet{cornelius_zno_2016} also observed a decrease in the lattice parameter with increasing substrate temperature.
However, they observed a slight increase after a minimum at a substrate temperature that depended on the Al concentration.

\begin{figure}
    \centering
    \includegraphics[width=0.5\textwidth]{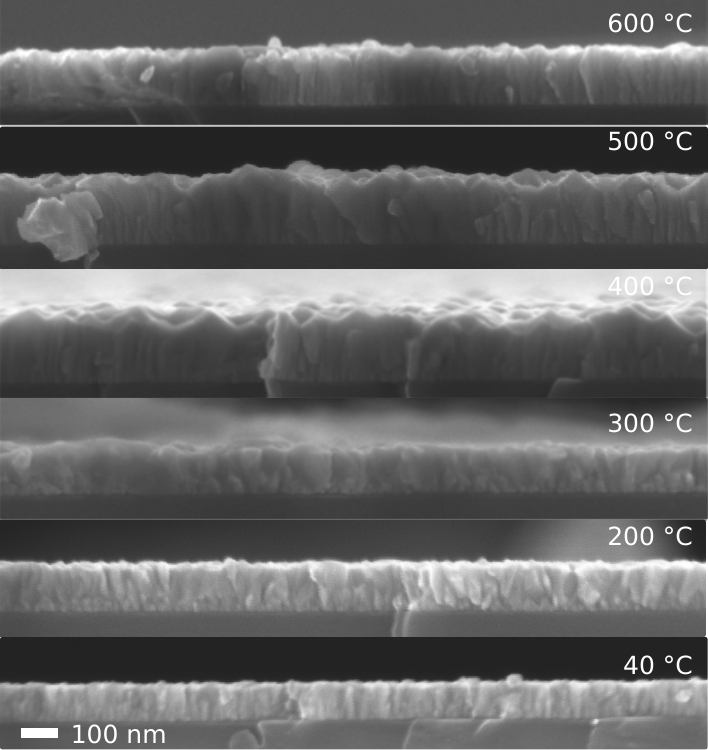}
    \caption{\label{fig:SEM-images}Cross-sectional SEM images of the AZO films from sample series~2 as a function of substrate temperature}
\end{figure}

Further information about the microstructure has been obtained from cross-sectional SEM images.
The SEM images of the films from sample series~2 are shown in figure~\ref{fig:SEM-images}.
The films appear to have grown in dense columns and without voids for all the substrate temperatures.
The lateral column size seems to be between 20\,nm and 50\,nm, while the larger columns are mostly found for the higher deposition temperatures.

\section{Discussion}

It is surprising that the deposition rate of the films is increasing with increasing substrate temperature up to 500\,$^\circ$C.
In theory, one would expect the thickness to decrease, as the sticking coefficient of the sputtered material on the substrate is expected to decrease with temperature.
Additionally, a preferential desorption of Zn is commonly observed at higher substrate temperatures together with an increase in the Al concentration due to the high vapor pressure of Zn~\cite{vinnichenko_establishing_2010}.
However, our EDS results do not show an increase in the Al concentration with the substrate temperature, meaning that evaporation processes are unlikely to play a large role in our deposition process.

The electrical properties of the AZO films are improving with the substrate temperature.
This improvement is mainly due to the increase in the mobility as can be seen in figure~\ref{fig:elec}b and, at least for series~2, also due to an increase in charge carrier concentration.
The increase in mobility can be related to the improvement in the crystallinity.
Both the intensity and the coherence length of the films are increasing up to 500\,$^\circ$C as can be seen in figures \ref{fig:XRD-combined}a and \ref{fig:XRD-combined}b, respectively.

Contrary to many reports in the literature~\cite{vinnichenko_establishing_2010,bikowski_influence_2012,cornelius_zno_2016}, we do not observe an optimum resistivity at an intermediate temperature of 200-350\,$^\circ$C.
\citet{bikowski_impact_2013} attribute the improvement in resistivity at intermediate temperatures to the thermal annealing of acceptor defects that are introduced by negative oxygen ion bombardment.
They explain the increase of resistivity at high temperature by the formation of secondary phases.
In our case, we do not observe an increase in the Al concentration with the temperature nor a decrease in the charge carrier concentration suggesting a small influence of oxygen ion bombardment.

In the case of HiPIMS, higher deposition temperatures can be considered in order to improve the electrical properties of AZO films.
As we have shown recently, the HiPIMS process enables a wider process window to synthesize transparent and conductive AZO films.
We related this to the possibility to maintain a low oxide coverage of the metallic target leading to a process taking place in the metal or transition mode~\cite{mickan_room_2016}.
The same argument can be used to interpret the present results.
It is rather well established that bombardment of the growing film by energetic oxygen ions is highly detrimental to the electrical properties of AZO by deactivation of the Al dopant, and introduction of oxygen interstitials~\cite{fumagalli_room_2014,bikowski_impact_2013}.
The formation of the homologous Al$_2$O$_3$(ZnO)$_m$ phase could also be related to the oxygen ion bombardment~\cite{horwat_deactivation_2010}.
Recent results have shown, that the formation of the homologous phase is less pronounced in HiPIMS processes~\cite{horwat_new_2016}.
In the present work, substoichiometric films were deposited at room temperature despite the high oxygen partial pressure.
The improved oxidation of the film at higher substrate temperatures can be related to the increased chemical reactivity.
Together with the improvement in the crystallinity due to thermal effects, the reduced oxygen ion bombardment due to the HiPIMS process allows the deposition of highly conductive, transparent AZO films up to substrate temperatures of 600\,$^\circ$C.

\section{Conclusion}

AZO films were deposited using reactive HiPIMS at substrate temperatures up to 600\,$^\circ$C.
The electrical properties were found to improve significantly with increasing substrate temperature with an optimum resistivity of $3 \times 10^{-4}$\,\ohm cm.
Achieving the best electrical properties at high deposition temperatures of 500 or 600\,$^\circ$C is in contrast to the many reports in literature about RF and reactive DC sputtering of AZO films, that find a degradation of the electrical properties at high substrate temperatures.
The good electrical properties can be related to the HiPIMS process, that allows the target to stay in the metal or transition mode even at high oxygen partial pressures.
Finally, the improvement of the crystalline properties due to the thermal effects leads to an improvement in the mobility of the charge carriers and an improved capability of the films to withstand ambient moisture.

\section*{Acknowledgements}
All authors thank Robert Boyd (Linköping University) for the help with the EDS measurements.
MM thanks the European Commission for the 'Erasmus Mundus' scholarship within the DocMASE project.

\section*{Supplementary material}
\label{suppl}

\section*{References}
\bibliography{references}

\end{document}